\documentclass[aps,pra,eqsecnum,floatfix,showpacs,twocolumn,tightenlines,
superscriptaddress]{revtex4}

\usepackage{graphicx}
\newcommand{\bra}[1]{\langle{#1}|}
\newcommand{\ket}[1]{|{#1}\rangle}
\newcommand{\tr}{{\rm tr \thinspace}}
\def\ketc[#1]{\vert #1 \rangle}
\def\brac[#1]{\langle #1 \vert}

\newcommand{\expect}[1]{\langle{#1}\rangle}
\newcommand{\expectc}[1]{\langle{#1}\rangle^{\phantom{\dagger}}_{\!c}}

\newcommand{\beq}{\begin{equation}}
\newcommand{\eeq}{\end{equation}}
\newcommand{\bqa}{\begin{eqnarray}}
\newcommand{\eqa}{\end{eqnarray}}
\newcommand{\nn}{\nonumber}

\newcommand{\erf}[1]{Eq.~(\ref{#1})}
\newcommand{\dg}{^\dagger}

\newcommand{\overt}{}
\begin{document}
\title{Quantum error correction for continuously detected 
errors}
\author{Charlene Ahn}
\affiliation{Institute for Quantum Information,
California Institute of Technology,
Pasadena, CA 91125, USA}
\email{cahn@theory.caltech.edu}
\author{H. M. Wiseman}
\affiliation{Centre for Quantum Computer Technology,  
Centre for Quantum Dynamics, School of Science, Griffith University,
Brisbane, QLD 4111 Australia}
\email{H.Wiseman@griffith.edu.au}
\author{G. J. Milburn}
\affiliation{Centre for Quantum Computer Technology, School of 
Physical
Sciences, The University of Queensland, QLD 4072 Australia}
\email{Milburn@physics.uq.edu.au}
\begin{abstract}
We show that quantum feedback control can be used as a quantum error
correction process for errors induced by weak continuous measurement. In
particular, when the error model is restricted to one, perfectly
measured, error channel per physical qubit, quantum feedback can act to
perfectly protect a stabilizer codespace. Using the stabilizer formalism
we derive an explicit scheme, involving feedback and an additional
constant Hamiltonian, to protect an ($n-1$)-qubit logical state encoded
in $n$ physical qubits. This works for both Poisson (jump) and
white-noise (diffusion) measurement processes. In addition, universal
quantum computation is possible in this scheme.  As an example, we show
that detected-spontaneous emission error correction with a driving
Hamiltonian can greatly reduce the amount of redundancy required to
protect a state from that which has been previously postulated [e.g.,
Alber \emph{et al.}, Phys. Rev. Lett. 86, 4402 (2001)].
\end{abstract}
\maketitle
\section{Introduction}

Many of the applications of quantum information science, such as quantum
computation \cite{ph219_notes,nc} and quantum cryptography
\cite{secure_QKD}, rely on preserving the coherence of quantum states.
However, these states are typically short-lived because of unavoidable
interactions with the environment. Combatting this decoherence has been
the subject of much study.

Two important tools that have been developed for this task are quantum
error correction  \cite{shor-ec,steane-ec,knill-laflamme,gott-stab}
and quantum feedback  \cite{wiseman_pra94,doherty-jacobs,WisManWang02}.
In the usual protocol for quantum error correction, projective
measurements are performed to acquire an error syndrome. A unitary
operation chosen based on the results of the projective measurements is
then applied to correct for the error. Quantum feedback control, on the
other hand, uses the tools of continuous measurements and Hamiltonian
feedback. The parameter to be controlled is typically the strength of
the feedback Hamiltonian, which is conditioned on the result of the
continuous measurements.

Quantum error correction and quantum feedback both rely on performing
operations that are conditioned on the result of some measurement on 
the
system, which suggests that exploring the links between these two
techniques adds to our understanding of both processes, and may lead 
to insights into future protocols and experimental implementations.  
In particular, this work provides an alternate avenue for examining the
situation considered in
\cite{mabuchi-zoller,qec-spont,detected-jump1,detected-jump2} of
correcting for a specific error process, such as spontaneous emission,
at the expense of correcting fewer general errors. Practically, as 
these authors
point out, it makes sense to pursue the tradeoff between general
correction ability and redudancy of coding, as smaller codes are more
likely to be in the range of what can be experimentally realized in 
the near future. We shall see that combining the pictures of quantum
feedback and error correction provides a convenient framework in which
to investigate this situation.

An additional motivation for considering the union of these 
techniques,
as in \cite{ADL}, is to examine what is possible with different 
physical
tools: in particular, continuous measurements and
Hamiltonians instead of the projective measurements and fast unitary
gates generally assumed by discrete quantum error correction. 
Continuous
error correction might well be useful even in a scenario in which
near-projective measurements are possible (e.g., ion
traps \cite{Wineland} and superconducting qubits \cite{Vion2002}); it
could be modified to provide bounds on how strong interactions in such
systems would have to be to perform operations such as error 
correction
and stay within a certain error threshold.

Ref.~\cite{ADL} presupposed that classical processing of currents could
be done arbitrarily quickly, so the feedback was allowed to be an
extremely complicated function of the entire measurement record. This
can be modeled only by numerical simulations. In this paper, by
contrast, we will restrict our feedback to be directly proportional to
measured currents, thus removing any need for classical
post-processing. In the Markovian limit, this allows an analytical
treatment.  This simplification is possible because in this paper we
assume that the errors are {\em detected}. That is, the experimenter
knows precisely what sort of error has occurred because the environment
that caused the errors is being continuously measured.  Since the
environment is thus acting as part of the measurement apparatus, the
errors it produces could be considered measurement-induced errors.

There are a number of implementations in which measurement-induced
 errors of this sort may be significant. In the efficient linear optics
 scheme of Knill {\em et al.} \cite{KLM}, gates are implemented by
 nondeterministic teleportation. Failure of the teleportation
 corresponds to a gate error in which one of the qubits is measured in
 the computational basis with known result. In a number of solid state
 schemes, the readout device is always present and might make an
 accidental measurement of a qubit, even if the readout apparatus is in
 a quiescent state. An example is the use of RF single electron
 transistors to readout a charge transfer event in the Kane
 proposal. Such a measurement is modelled as a weak continuous
 measurement \cite{Wiseman2001}. While one supposes that the SET is
 biased in its low conductance state during qubit processing, it is
 useful to know that even if the device does accidentally make a
 measurement, the resulting error can be corrected.

In this paper, we show that for certain error models and codes,
Markovian feedback plus an additional constant Hamiltonian (a ``driving
Hamiltonian'') can protect an unknown quantum state encoded in a
particular codespace.  Using the stabilizer formalism, we show that if
there is one sort of error per physical qubit, and the error is detected
perfectly, then it is always possible to store $n-1$ logical qubits in
$n$ physical qubits. This works whether the detector record consists of
discrete spikes (Poisson noise) or a continuous current (white
noise). This suggest that if the dominant decoherence process can be
monitored, then using that information to control the system Hamiltonian
may be the key to preventing such decoherence (see also the example in
\cite{doherty_thesis}).

As a salient application of this formalism, we consider the special case
of spontaneous emission. Stabilizing states against spontaneous emission
by using error-correcting codes has been studied by several groups
\cite{mabuchi-zoller,qec-spont,detected-jump1,detected-jump2}.  Here we
demonstrate that a simple $n$-qubit error-correcting code, Markovian
quantum feedback, and a driving Hamiltonian, is sufficient to correct
spontaneous emissions on $n-1$ qubits.  The result of encoding $n-1$
logical qubits in $n$ physical qubits has been recently independently
derived by \cite{K-Lidar} for the special case of spontaneous emission;
however, our scheme differs in a number of respects. We also show that
spontaneous emission error correction by feedback can be incorporated
within the framework of canonical quantum error correction, which can
correct arbitrary errors.

The paper is organized as follows. We review some useful results in
quantum error correction and quantum feedback theory in Sec.\
\ref{sec:background}. In Sec.\ \ref{sec:ham-app} we present the example
of detected spontaneous-emission errors, first for 2 qubits and then for
$n$ qubits.  In Sec.\ \ref{sec:gen} we generalize this for protecting an
unknown state subject to any single-qubit measurements. We show how to
perform universal quantum computation using our protocol in Sec.\
\ref{sec:uqc}. Sec.\ \ref{sec:conclusion} concludes.

\section{Background}\label{sec:background}

\subsection{Quantum error correction: stabilizer 
formalism}\label{bgqecsf}

Quantum error correction has specifically been designed for protecting
unknown quantum states \cite{gott-stab,knill-laflamme,ftqc}. An
important class of quantum error-correcting codes are the stabilizer
codes. An elegant and simple formalism \cite{gott-stab} exists for
understanding these codes; in this paper we will restrict ourselves to
this class of codes in order to take advantage of this formalism.

In the remainder of this paper we will use the notation of
\cite{nc} in which $X$, $Y$, and $Z$ denote the Pauli 
matrices
$\sigma_x$, $\sigma_y$ and $\sigma_z$ respectively, and juxtaposition
denotes a tensor product; hence any element of the Pauli group
\begin{equation}
P_n = \{\pm 1, \pm i\}\otimes \{I,X,Y,Z\}^{\otimes n}
\end{equation}
may be denoted as a concatenation of letters (e.g., $ZZI = \sigma_z
\otimes \sigma_z \otimes I$).

A stabilizer code may be defined simply as follows: Consider a
$2^n$-dimensional ($n$-qubit) Hilbert space and a subgroup of 
$2^{n-k}$
commuting Pauli operators ${\cal S} \in P_n$. This group of operators 
is
the \emph{stabilizer} of the code; the codespace $\cal C(S)$ is the
simultaneous $+1$ eigenspace of all the operators in $\cal S$. It can 
be
shown that if $-I$ is not an element of ${\cal S}$, the subspace
stabilized is non-trivial, and the dimension of $\cal C(S)$ is 
$2^{k}$;
hence, we regard this system as encoding $k$ qubits in $n$.
The generators of such a group are a subset of this group such
that any element of the stabilizer can be described as a product of
generators. It is not hard to show that $n-k$ generators suffice to
describe the stabilizer group ${\cal S}$.

When considering universal quantum computation it is also useful to
define the {\em normalizer} of a code. Given a stabilizer group ${\cal
S}$, the normalizer $N({\cal S})$ is the group of elements in $P_n$ that
commute with all the the elements of ${\cal S}$, and it can be shown
that the number of elements in $N({\cal S})$ is $2^{n+k}$.

Now, $n + k$ generators suffice to describe $N({\cal S})$. Of these, $n
- k$ can be chosen to be the generators of ${\cal S}$. It can be shown
that the remaining $2 k$ generators can be chosen to be the {\em encoded
operators} $\bar{Z}_\mu, \bar{X}_\mu, \mu = 1,2,...,k$, where $\bar{Z}_\mu,
\bar{X}_\mu$ denote the Pauli operators $X$ and $Z$ acting on encoded
qubit $\mu$, tensored with the identity acting on all other encoded
qubits. These encoded operators act, as their name implies, to take
states in $\cal C(S)$ to other states in $\cal C(S)$.

The usual protocol for stabilizer 
codes,
which will be modified in what follows, starts with measuring the stabilizer
generators. This projection discretizes whatever error has occurred 
into
one of $2^{n-k}$ error syndromes labeled by the $2^{n-k}$ possible
outcomes of the stabilizer generator measurements. The information 
given
by the stabilizer measurements about what error syndrome has occurred 
is
then used to apply a unitary recovery operator that returns the state 
to
the codespace.

\begin{figure}[h!]
\includegraphics[width=0.4\textwidth]{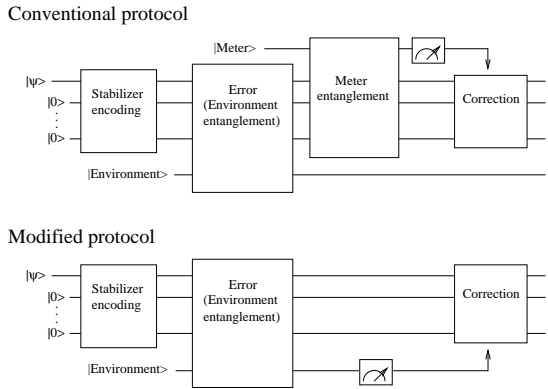}
\caption{The top diagram shows the conventional stabilizer error
correction protocol. After the state is encoded, an error occurs through
coupling with the environment. To correct this error, the encoded state
is entangled with a meter in order to measure the stabilizer generators,
and then feedback is applied on the basis of those measurements. The
bottom diagram shows our modified protocol, in which the error and
measurement steps are the same. To correct the error in this protocol,
the environment qubits are measured, and we feedback on the results of
the environment measurement. } \label{fig:ecfig}
\end{figure}

In this paper we will use a modified version of this protocol. In
particular, we will not measure stabilizer elements.  Instead, we will
assume that a limited class of errors occurs on the system and that
these errors are detectable: we know when an error has happened and 
what
the error is. The correction back to the codespace can still be
performed by a unitary recovery operator based on the information from
the error measurement. Fig.\ \ref{fig:ecfig} shows the difference
between the conventional protocol and our modified protocol.

In this paper, we will also consider operators of the form
\begin{equation}
T = T_1 \otimes \cdots \otimes T_n,
\end{equation}
where $T_i$ is an arbitrary traceless one-qubit operator normalized such
that its eigenvalues are $\{-1,1\}$. Operators of this form are not
generally Pauli-group stabilizers as presented in \cite{gott-stab}, as
$T$ is not in general a member of $P_n$. However, because of the special
form of $T$, $T$ is equivalent to a Pauli operator up to conjugation by
a unitary that is a product of one-qubit unitaries, i.e., there exists
some $U = \bigotimes_{i=1}^{n} U_i$ such that $U T U^\dagger$ is a
member of $P_n$.  Therefore, choosing $T$ as the sole stabilizer
generator for a code is equivalent, up to conjugation by a unitary, to
choosing a member of the Pauli group as the stabilizer generator.  (Note
that additional constraints are necessary if $T$ is not the only
stabilizer generator.)

\subsection{Quantum feedback}\label{sec:ql-fb}

Continuous quantum feedback can be defined, for the present purposes, as
the process of monitoring a quantum system and using the continuous (in
time) measurement record to control its dynamics. It can be analysed by
considering the dynamics of the measured system conditioned on the
continuous measurement record; this process is referred to as {\em
unraveling}. The reduced dynamics of a system subject to weak
continuous measurement is described by a Markov master equation, which
determines the dynamics of the system averaged over all possible
measurement records. However, if the time-continuous measurement record
(a classical stochastic process) is known, then it is possible to
describe the conditional state of the measured system by a stochastic
conditional evolution equation.  A given master equation does not
uniquely determine the conditional evolution equation, as there are many
ways in which information about the system may be collected from the
environment to which it is coupled as a result of the measurement.  That
is to say, a given master equation admits many unravelings.

In this section we will introduce some of the results of this 
formalism; for more details see \cite{wiseman-semiclass}.  We will assume
that the change in the state of the system over a time interval $dt$
due to its interaction with the environment can be described by a 
single jump operator $c$. By this we mean that 
jumps are  represented by a Kraus operator $\Omega_1 = c
\sqrt{dt}$, so that they occur with probability $\expect{c^\dagger c}
dt$.  Normalization requires another Kraus operator,
$\Omega_0 = 1 - c^\dagger c dt / 2 - i H dt$, where $H$ is 
Hermitian. 
Then the unconditional
master equation without feedback is just the familiar Lindblad form
\cite{carmichael}
\begin{eqnarray}
\label{eqn:lindblad}
d\rho &=& \Omega_0 \rho \Omega_0
  + \Omega_1 \rho \Omega_1 - \rho\nonumber\\
&=& -i[H,\rho]dt + c \rho c^\dagger dt 
 - \frac{1}{2} (c^\dagger c \rho + \rho c^\dagger c) dt\nonumber\\
&\equiv&  -i[H,\rho]dt + {\cal D} [c] \rho dt.
\end{eqnarray}
A bosonic example is given in \cite{WallsMilb}, while 
a fermionic example is given in \cite{Milburn1998}.

\subsubsection{Jump unravelings} 

One way to 
unravel this master equation is to assume that the environment is 
measured so that the time of each 
jump event is determined. If the measured 
number of jumps up to time $t$ is denoted $N(t)$, then 
the increment $dN(t)$ is defined by 
\begin{eqnarray}
\label{eqn:dN1}
dN_c(t)^2 & = & dN_c(t)\\
\label{eqn:dN2}
E[dN_c(t)] & = & 
\expect{c^\dagger c}_c dt.
\end{eqnarray}
Here $E[\  ]$ defines a 
classical ensemble average, and the subscript $c$ on the quantum 
average reminds us that the rate of the process at time $t$ depends 
on the conditional state of the quantum system up to that time. That 
is to say, it depends on the state of the quantum system conditioned on the 
entire previous history of the current $dN/dt$. This conditional state is 
determined by a
stochastic Schr\"{o}dinger 
equation
\begin{eqnarray}
\label{eqn:SSE-jump}
d|\psi_c(t)\rangle &=& \left[dN_c(t)\left(\frac{c}{\sqrt{\langle
c^\dagger c\rangle_c(t)}}-1\right) + dt \right. \nn \\
&&\times \,\left. \left(\frac{\langle c^\dagger
c\rangle_c(t)}{2}-\frac{c^\dagger c}{2}-iH\right)\right]|\psi_c(t)\rangle .
\end{eqnarray}
We will refer to this as a {\em jump unraveling}. If we average over the
measurement record to form
$\rho(t)=E[|\psi_c(t)\rangle\langle\psi_c(t)|]$, it is easy to show
using Eqs. (\ref{eqn:dN1}) and (\ref{eqn:dN2}) that $\rho(t)$ obeys the
unconditional master equation given in Eq.(\ref{eqn:lindblad}).

Now consider Markovian Hamiltonian feedback, linear in the current:
\begin{equation}
H_{fb}(t) = \frac{dN(t)}{dt} V,
\end{equation}
with $V$ an Hermitian operator. Taking into account that the feedback
must act after the measurement, it can be shown \cite{wiseman_pra94}
that the feedback modifies the conditional evolution by changing the $c$
in the numerator of the first term into $e^{-iV}c$. Since likewise
changing all of the other occurrences of $c$ has no effect, the ensemble
average behaviour is the same as before, with $c$ changed to
$e^{-iV}c$. That is to say, the feedback-modified master equation is
\begin{equation}
\dot{\rho} = -i[H,\rho] + {\cal D}[e^{-iV}c]\rho.
\end{equation}

\subsubsection{Diffusive unravelings}

A very different unraveling may be defined by first noting that  
given
some complex number $\gamma = \vert \gamma \vert e^{i \phi}$, we
may make the transformation
\begin{eqnarray}
\label{eqn:xform}
c &\rightarrow& c + \gamma\nonumber\\
H &\rightarrow& H -\frac{i \vert \gamma \vert}{2} 
   ( e^{-i \phi} c -  e^{i \phi} c^\dagger)
\end{eqnarray}
and obtain the same master equation. In the limit as $\vert \gamma
\vert$ becomes very large, the rate of the Poisson process is dominated
by the term $|\gamma|^2$.  In this case it may become impossible to
monitor every jump process, and a better strategy is to approximate 
the Poisson stochastic process by a Gaussian white-noise process.

For large $\gamma$, we can consider the system for a time $\delta t$ 
in which the system changes negligibly but the number of detections 
$\delta
N(t) \approx \vert \gamma \vert ^2 \delta t$ is very large; then we 
can
approximate $\delta N(t)$ as \cite{fieldquad}
\begin{equation}
\delta N(t) \approx \vert \gamma \vert^2 \delta t  + \vert \gamma \vert
\expectc{e^{-i\phi}c + c^{\dagger}e^{i\phi}}\delta t  + \vert \gamma \vert
\delta W(t),
\end{equation}
where $\delta W(t)$ is normally distributed with mean zero and 
variance
$\delta t$.

We now define the stochastic measurement record as the current
\begin{eqnarray}
\label{eqn:photocurrent}
\frac{dQ(t)}{dt} &=& \lim_{\gamma \rightarrow \infty} \frac{\delta N(t) 
-   \vert \gamma \vert^2 \delta t}{\vert \gamma \vert \delta t} \\
&=& \expectc{ e^{-i \phi}c +  e^{i \phi}c^\dagger}   
 + dW(t)/dt. \label{eqn:current}
\end{eqnarray}
Given this stochastic measurement record, 
we can determine the conditional state 
of the quantum system by a stochastic  Schr\"{o}dinger 
equation analogous to Eq. (\ref{eqn:SSE-jump}). The equivalence (in the
ensemble average) to the master equation (\ref{eqn:lindblad}) is, in
this case, easier to see by considering $\rho_c = \ket{\psi_c}\bra{\psi_c}$,
which obeys the stochastic master equation
\begin{eqnarray}
\label{eqn:sme} d\rho_c(t) &=& -i\left[H,\,\rho_c(t)\right] dt 
+  {\cal D}[e^{-i \phi} c] \rho_c(t) dt 
\nonumber\\  & &   + \, {\cal H}[e^{-i \phi} c] \rho_c(t) dW(t). 
                               \label{SME}
\end{eqnarray}
In the above equations, the expectation $\expectc{a}$ denotes
$\tr(\rho_c a)$, $dW$ is a normally
distributed infinitesimal random variable with mean zero and variance
$dt$ (a \emph{Wiener increment} \cite{gardiner}), and ${\cal H}$ is a
superoperator that takes a jump operator as an argument and acts on
density matrices as
\begin{eqnarray}
{\cal H}[c]\rho &=& c\rho + \rho c^\dagger
                   - \rho\, \tr[{c\rho + \rho c^\dagger}].
\end{eqnarray}

We thus have a different unraveling of the original master equation
Eq.(\ref{eqn:lindblad}). Because of the white noise in the stochastic
master equation (\ref{SME}) we call this a diffusive unraveling.  It
applies, for example, when one performs a continuous weak homodyne
measurement of a field $c$ by first mixing it with a classical local
oscillator in a beamsplitter and then measuring the output beams with
photodetectors \cite{fieldquad}. In that case the measurement process
$dQ(t)$ determines the observed photocurent.  Another measurement model
in which it may be appropriate to approximate a Poisson measurement
process by a white-noise measurement process is the electronic point
contact model for monitoring a single quantum
dot\cite{Goan2001b,Goan2001}. In that case the form of the master
equation itself determines a large background jump rate, rather than an
imposed classical field prior to detection.

We now consider Markovian feedback of the white-noise measurement 
record via a
Hamiltonian, where the strength of the feedback is a linear function 
of the
measurement current:
\begin{equation}
H_{fb}(t) = \frac{dQ(t)}{dt} F,
\end{equation}
where $F$ is a Hermitian operator. 
It can be shown that the addition of
such feedback leads to the conditioned master equation 
\cite{wiseman_pra94,WisMilprl93}
\begin{eqnarray}
\dot{\rho} &=&  -i [( e^{i \phi}c^\dagger F +  e^{-i \phi} F c)/2
 +H, 
\rho] \nn\\ && 
 \, + {\cal D}[ e^{-i \phi}c - i F] \rho \nn\\ && 
\,+ dW(t) {\cal H}[e^{-i \phi}c - i F]\rho.
\end{eqnarray}
In order to derive analytic results given such feedback, it is
convenient to consider the average over many such evolution
trajectories. Since the expectation value of $dW$ is zero, averaging
yields an unconditioned master equation 
\begin{eqnarray}
\dot{\rho} &=&  -i [( e^{i \phi}c^\dagger F +  e^{-i \phi} F c)/2
 +H, 
\rho]\nonumber\\ 
&&  + {\cal D}[ e^{-i \phi} c - i F] \rho \label{eqn:gen-fb-me}
\end{eqnarray}
Note that these equations are only valid for perfect (unit-efficiency)
detection; the correspondences between error correction and
feedback are more readily seen in this case, and we discuss the
case of imperfect detection in Sec.\ \ref{sec:discussion}.

These feedback equations are easily generalized in the following way: 
Given
$n$ qubits, denote a set of measurement operators by $\{c_1, c_2, \cdots,
c_n\}$, where $c_j$ acts on the $j$th qubit, and a set of feedback
operators by $\{F_1,\cdots, F_n\}$, where the action of $F_j$ is 
conditioned
on the measurement of the $j$th qubit. Then the unconditional master
equation (\ref{eqn:gen-fb-me}), for example, generalizes to
\begin{eqnarray}
\dot{\rho} = \sum_{j=1}^{n} && 
     \{ -i [( e^{i \phi_j} c_j^\dagger F_j +  e^{-i \phi_j} F_j 
c_j)/2  +H, \rho]\nonumber\\ 
    &&  + {\cal D}[ e^{i \phi_j} c_j - i F_j] \rho \}. 
    \label{eqn:lin-fb-me}
\end{eqnarray}

\section{Example: Spontaneous-emission correction}
\label{sec:ham-app}
A particular example of a Poisson process error is spontaneous 
emission,
in which the jump operator is  proportional to $|0\rangle\langle 1|$, 
so that the state
simply decays from $|1\rangle$ to $|0\rangle$ at random times.  
Indeed,
if the decay is observed (say by emitting a photon which is then
detected), this may be regarded as a destructive measurement of
the operator $|1\rangle\langle 1|$.

Stabilizing states against the important decay process of spontaneous
emission through application of error-correcting codes has been 
studied
by several groups
\cite{mabuchi-zoller,qec-spont,detected-jump1,detected-jump2}. In
\cite{qec-spont} Plenio, Vedral and Knight considered the structure of
quantum error correction codes and addressed the problem that 
spontaneous
emission implies continuous evolution of the state even when no 
emission
has occurred. They developed an eight-qubit code that both corrects one
general error and corrects the no-emission evolution to arbitrary 
order.

 More recently, in several papers Alber \emph{et al.}
\cite{detected-jump1,detected-jump2} have addressed a somewhat more
specific problem relating to spontaneous emission from statistically
independent resevoirs. In this formulation, the only errors possible 
are
spontaneous emission errors, and the time
and position of a particular spontaneous emission is known. They 
showed
that given these constraints, a reduction of the redundancy in
\cite{qec-spont} was possible, and constructed a four-qubit code which
corrects for one spontaneous emission error. 

Here we show that for the case considered in
\cite{detected-jump1,detected-jump2}, a very simple error correcting
code consisting of just two qubits with feedback  is
sufficient to correct spontaneous emissions  for a single logical qubit.  
A crucial difference 
from Refs.~\cite{detected-jump1,detected-jump2} is that we 
call for a constant driving Hamiltonian in addition to the feedback 
Hamiltonian. Moreover, a simple code of
$n$ qubits, with the appropriate feedback and driving Hamiltonians,
can encode $n-1$
qubits and correct for spontaneous emissions when the position (i.e. 
which qubit) and time of the jump are known.  
We also show that an equally effective protocol can be found 
for a diffusive unraveling of the spontaneous emission (as in 
homodyne detection).

\subsection{Two-qubit code: Jump unraveling}
\label{subsec:2qju}

The simplest system for which we can protect against detected
spontaneous emissions is a system of two qubits.  We consider the model
in which the only decoherence process is due to spontaneous emission
from statistically independent reservoirs. We will show that a simple
code, used in conjunction with a driving Hamiltonian, protects the
codespace when the time and location of a spontaneous emission is known and a
correcting unitary is applied instantaneously; the codespace suffers no
decoherence.

The codewords of the code are given by the following:
\begin{eqnarray}
\label{eqn:cw}
\ket{\bar{0}} &\equiv& (\ket{00} + \ket{11})/\sqrt{2} \nonumber\\
\ket{\bar{1}} &\equiv& (\ket{01} + \ket{10})/\sqrt{2}. 
\end{eqnarray}
In the stabilizer notation, this is a stabilizer code with
stabilizer generator $XX$.
Both codewords are $+1$ eigenstates of $XX$.

Following the presentation in Sec.~\ref{sec:ql-fb}, the jump 
operators for spontaneous emission of the $j$th qubit are
\begin{equation}
\Omega_{j} = \sqrt{\kappa_j dt} (X_{j} - i Y_{j}) \equiv \sqrt{\kappa_j dt}a_j,
\end{equation}
where $4 \kappa_{j}$ is the decay rate for that qubit. 
In the absence of any feedback, the master equation is
\begin{equation}
\dot{\rho} = \sum_{j=1,2} \kappa_{j}{\cal D}[X_{j} - i Y_{j}]\rho 
-i[H,\rho]. \label{spemme}
\end{equation}

If the emission 
is detected, such that the qubit $j$ from which it originated is known, 
it is possible to correct back to the codespace without knowing the 
state. This is because 
the code and error fulfill the necessary and sufficient
conditions for appropriate recovery operations \cite{knill-laflamme}:
\begin{equation}
\label{eqn:kl-cond}
\bra{\psi_\mu} E^ \dagger E \ket{\psi_\nu} = \Lambda_E \delta_{\mu\nu}.
\end{equation}
Here $E$ is the operator for the 
measurement (error) that has  occurred  and $\Lambda_{E}$ is 
a constant. The states 
 $\ket{\psi_\mu}, \ket{\psi_\nu}$ are the encoded states in 
 Eq.~(\ref{eqn:cw}) with $\langle 
\psi_\mu
\vert \psi_\nu \rangle = \delta_{\mu\nu}$.  These conditions differ from the
usual condition only by taking into account 
that we \emph{know} a particular error 
$E=\Omega_{j}$ has occurred.

More explicitly, if a spontaneous emission on the first qubit occurs,
$\ket{\bar{0}} \rightarrow \ket{01}$ and $\ket{\bar{1}} \rightarrow
\ket{00}$, and similarly for spontaneous emission on the second
qubit. Since these are orthogonal states, 
this fulfills the condition given in (\ref{eqn:kl-cond}), so a
unitary exists that will correct this spontaneous emission error. 
One choice for the correcting unitary is
\begin{eqnarray}
\label{eqn:spont-em-U}
U_1 &=& (XI - ZX)/\sqrt{2} \nonumber\\
U_2 &=& (IX - XZ)/\sqrt{2}.
\end{eqnarray}

As pointed out in \cite{qec-spont}, a further complication is the
nontrivial evolution of the state in the time between spontaneous
emissions. From Sec.~\ref{sec:ql-fb}, this is described by the 
measurement operator 
\begin{eqnarray}
\label{eqn:omega0-twobit}
\Omega_0 &=& II\left(1 - (\kappa_1 + \kappa_2) dt\right) - 
 {\kappa_1 dt} ZI  \nonumber\\ 
&& - {\kappa_2 dt}  IZ   - i Hdt.
\end{eqnarray}
The non-unitary part of this evolution 
can be  corrected by assuming a driving 
Hamiltonian of the form
\begin{equation}
\label{eqn:spont-driving}
H =-(\kappa_1 YX  +\kappa_2 XY).
\end{equation}
This result can easily be seen by plugging (\ref{eqn:spont-driving}) 
into
(\ref{eqn:omega0-twobit}) with a suitable rearrangement of terms:
\begin{eqnarray}
\label{eqn:omega0-new}
\Omega_0 &=& II\left(1 - 
(\kappa_1 + \kappa_2) dt\right) -  {\kappa_1 dt}  ZI (II - 
XX) \nonumber \\
&& - \, {\kappa_2 dt}  IZ (II - XX),
\end{eqnarray}
and since $II - XX$ acts to annihilate the codespace, $\Omega_0$ acts
trivially on the codespace.

 We then have the following master
equation for the evolution of the system:
\begin{equation}
\label{eqn:twobit-ec-me}
d{\rho} = \Omega_0 \rho \Omega^{\dagger}_0 - \rho + dt \sum_{j = \{1,2\}}
\kappa_{j} U_{j} a_{j} \rho a^{\dagger}_j U^\dagger_{j}, 
\end{equation}
where $U_j$ is the recovery operator for a spontaneous emission from
qubit $j$. From Sec.~\ref{sec:ql-fb}, these unitaries can be achieved 
by the feedback Hamiltonian
\begin{equation}
H_{fb} = \sum_{j=1,2}\frac{dN_{j}(t)}{dt} V_{j},
\end{equation}
where $N_{j}(t)$ is the spontaneous emission count for qubit $j$, 
and $U_{j} = \exp(-iV_{j})$.  Here, we can see from the simple form of
(\ref{eqn:spont-em-U}) that $V_j$ can be chosen as proportional to  $U_j$.
Since $U_{j} a_{j} \rho a^{\dagger}_j U^\dagger_{j}$  
acts as the identity on the codespace by definition, and since we have 
shown that $ \Omega_0 \rho \Omega^{\dagger}_0$ preserves the codespace,
(\ref{eqn:twobit-ec-me}) must preserve the codespace.

Such a code is optimal in the sense that it uses the
smallest possible number of qubits required to perform the task of
correcting a spontaneous emission error, as we know that the
information stored in one unencoded qubit is destroyed by spontaneous
emission.

\subsection{Two-qubit code: Diffusive unraveling}
\label{subsec:2qcdu}
A similar situation applies for feedback of a continuous measurement
record with white noise, as from homodyne detection of the emission.  We
use the same codewords, and choose $\phi_j = -\pi/2$ for the
measurement.   Then (\ref{eqn:spont-em-U}) suggests using the
following feedback operators:
\begin{eqnarray}
F_1 &=& \sqrt{\kappa_1} (XI - ZX)  \nonumber\\
F_2 &=& \sqrt{\kappa_2} (IX - XZ).
\end{eqnarray}
 If we use these feedback Hamiltonians  with the same driving
Hamiltonian (\ref{eqn:spont-driving}) as in the jump case, the
resulting master equation is, using (\ref{eqn:lin-fb-me}),
\begin{eqnarray}
\label{eqn:markov-fb}
\dot{\rho} &=& \kappa_1{\cal D}[ YI - i ZX]\rho + \kappa_2{\cal D}[IY - i XZ ]\rho 
\end{eqnarray}

We can see that this master equation preserves the codespace, by again
noting that $YI - i ZX = YI (II - XX)$, and similarly for $IY - i
XZ$. The operator $II - XX$ of course acts to annihilate the
codespace. This insight will be used in the next section to derive a
feedback procedure for a more general measurement operator.

\subsection{Generalizations to $n$ qubits}

We will now demonstrate a simple $n$-qubit code that corrects for
spontaneous emission errors only, while encoding $n-1$ qubits. Both of 
the above calculations (jump and diffusion) 
generalize. The master equation is the same as (\ref{spemme}), but now the 
sum runs from 1 to $n$. Again we need only a single stabilizer 
generator, namely $X^{\otimes n}$. The number of codewords is thus 
$2^{n-1}$, enabling $n-1$ logical qubits to be encoded.
 Since it uses only one   physical qubit in excess of the number of 
 logical qubits, 
this is again  obviously an optimal code.

First, we consider the jump case. 
As in Sec.~\ref{subsec:2qju}, a spontaneous emission jump fulfills the
error-correction condition (\ref{eqn:kl-cond}) (see
Sec. \ref{sec:gen-unrav} below).  Therefore, there 
exists
a unitary that will correct for the spontaneous-emission
jump. Additionally, it is easy to see by analogy with
(\ref{eqn:omega0-new}) that
\begin{equation}
\label{eqn:n-spont-H}
H = \kappa_j \sum_j X^{\otimes j-1} Y X^{\otimes n-j} 
\end{equation}
protects against the nontrivial no-emission evolution.
 Therefore the codespace is protected. 

Next, for a diffusive unraveling, we again choose $\phi_{j} = -\pi/2$, 
as in Sec.~\ref{subsec:2qcdu}. The same driving Hamiltonian 
(\ref{eqn:n-spont-H}) is again required, and the 
feedback operators generalize to 
\begin{equation}
F_j = \sqrt{\kappa_j} \left(I^{\otimes j-1} X I^{\otimes n-j} +
 X^{\otimes j-1} Z X^{\otimes n-j}\right).
\end{equation}
The master equation becomes
\begin{equation}
\dot{\rho} = \sum_{j} \kappa_{j} {\cal D}[I^{\otimes j-1}YI^{\otimes 
n-j} (I^{\otimes n}-X^{\otimes n})].
\end{equation}

These schemes with a driving Hamiltonian do 
not have the admittedly desirable
property of the codes given in
\cite{qec-spont,detected-jump1,detected-jump2} that if there is a time
delay between the occurrence of the error and the application of the
correction, the effective no-emission evolution does not lead to
additional errors. Nevertheless, as pointed out in
\cite{detected-jump2}, the time delay for those codes must still be
short so as to prevent two successive spontaneous emissions between
correction; they numerically show that the fidelity decays roughly
exponentially as a function of delay time.  Therefore, we believe that
this drawback of our protocol is not significant.

\section{One-qubit general measurement operators}\label{sec:gen}
 
The form of the above example strongly indicates that there is a
nice generalization to be obtained by considering stabilizer 
generators in more detail. In this section, we consider 
an arbitrary measurement operator  operating on each 
qubit. We find the condition that the stabilizers of the codespace must 
satisfy. We show that it is always possible to find an optimal 
codespace (that is, one with a 
single stabilizer group generator). 
We work out the case of diffusive feedback in detail and
derive it as the limit of a jump process.

\subsection{General unraveling}\label{sec:gen-unrav}

Different unravelings of the master equation (\ref{eqn:lindblad}) may 
be usefully parameterized by $\gamma$. In Sec.\ \ref{sec:ql-fb}, we
have seen that a simple jump unraveling has $\gamma = 0$,
while the diffusive unraveling is characterized by $|\gamma| 
\rightarrow \infty$. 
We will now address the question of when a unitary correction
operator exists for arbitrary $\gamma$, i.e., when a measurement 
scheme with a given $\gamma$ works to correct the error.

Consider a Hilbert space of $n$ qubits with a stabilizer group 
$\{S_l\}$. Let us consider a single jump operator $c$ acting on  a
single qubit. We may then write $c$ in terms of Hermitian operators 
$A$ and $B$ as
\begin{eqnarray}
\label{eqn:decomp}
e^{-i \phi}c &=& \chi I + A + i B \\
&\equiv& \chi I + \vec{a} \cdot \vec{\sigma} + i \vec{b} \cdot 
\vec{\sigma}
\end{eqnarray}
where $\chi$ is a complex number,  $\vec{a}$ and $\vec{b}$ are real
vectors, and $\vec\sigma = (X,Y,Z)^{T}$. 

We now use the standard condition
(\ref{eqn:kl-cond}), where here we take $E = c + \gamma$. Henceforth, 
$\gamma$ is to be understood as real and positive, since the relevant phase 
$\phi$ has been taken into account in the definition (\ref{eqn:decomp}).
The relevant term is
\begin{eqnarray}
\label{eqn:D-def}
E^\dagger E &=& (|\chi + \gamma|^2 + \vec{a}^2 + \vec{b}^2) I 
\nonumber\\
&&+ \mathrm{Re}(\chi + \gamma) A + \mathrm{Im} (\chi + \gamma)^* i 
B + (\vec{a}\times
\vec{b}) \cdot \vec{\sigma})\nonumber\\
&\equiv& (|\chi + \gamma|^2 + \vec{a}^2 + \vec{b}^2) I + D,
\end{eqnarray}
where $D$ is Hermitian.

Now we can use the
familiar 
sufficient condition for a stabilizer code \cite{gott-stab}: the
stabilizer should anticommute with the traceless part of $E^\dagger
E$. This condition becomes
explicitly
\begin{equation}
\label{eqn:ec-gamma}
0 = \{S, D \}.
\end{equation}
As long as this is satisfied, there is some feedback unitary $e^{-iV}$ 
which will correct the error.

Normalization implies that when $E$ does not occur, there may still be
nontrivial evolution. In the continuous time paradigm, where one Kraus operator 
is
given by $E \sqrt{dt}$, the transform (\ref{eqn:xform}) tells us that
the no-jump normalization Kraus operator is given by
\begin{equation}
\label{eqn:omega0}
\Omega_0 = 1 -  \frac{1}{2} E^\dagger E dt -\frac{\overt \gamma \overt}{2} 
   ( e^{-i \phi} c -  e^{i \phi} c^\dagger) dt - i H dt.
\end{equation}
Now we choose the driving Hamiltonian 
\begin{equation}
\label{eqn:gen-ham-driv}
H = \frac{i}{2} D S +\frac{i \overt \gamma \overt}{2}(e^{-i \phi} c -   e^{i \phi}
c^\dagger).
\end{equation}
This is a Hermitian operator because of (\ref{eqn:ec-gamma}).
Then  the
total evolution due to $\Omega_0$ is   just the identity, apart from
a term
proportional to $D(1 - S)$, which annihilates the codespace.
Thus for a state initially in the codespace, the condition 
(\ref{eqn:ec-gamma})
suffices for correction of both the jump and no-jump evolution. 

A nice generalization may now be found for a set $\{ c_j \}$ of errors 
such
that $c_j$ [with associated operator $D_j$ as defined in
(\ref{eqn:D-def})] acts on the $j$th qubit alone.  Since $D_j$ is
traceless, it is always possible to find some other Hermitian 
traceless
one-qubit operator $s_j$ such that $\{ s_j, D_j \} = 0$. Then we may
pick the stabilizer group by choosing the single stabilizer generator
\begin{equation}
\label{eqn:oneS}
S = s_1 \otimes \cdots \otimes s_n
\end{equation}
so that the stabilizer group is $\{1, S\}$. As noted in 
Sec.~\ref{bgqecsf}, this is not strictly a stabilizer group, as $S$ 
may not be in the Pauli 
group, but this  does not change the analysis. 
Choosing $H$ according to this $S$ such 
that
\begin{equation}
\label{eqn:wn-ham-driv}
 H = \sum_j \frac{i}{2} D_j S +\frac{i \overt \gamma_{j} \overt}{2}(e^{-i \phi_j}
 c_j -   e^{i \phi_j} c_j^\dagger)
\end{equation}
 will, by
our analysis above, provide a total evolution that protects the
codespace, and the errors will be correctable; furthermore, this
codespace encodes $n-1$ qubits in $n$.

Note that we can now easily understand the $n$-qubit jump process
error of spontaneous emission considered in Sec.\
\ref{sec:ham-app}. Here, $\gamma = 0$, $S = X^{\otimes n}$, and $D_j =
2 \kappa_j Z_j$. Thus (\ref{eqn:ec-gamma}) is satisfied, and the
Hamiltonian (\ref{eqn:n-spont-H}) is derived directly from
(\ref{eqn:wn-ham-driv}).

Moreover, one is not restricted to the case of one stabilizer; it is
possible to choose a different $S_j$ for each individual error
$c_j$. For example, for the spontaneous emission errors $ c_j = X_j - i
Y_j$ we could choose $S_j$ as different stabilizers of the five-qubit
code. This choice is easily made, as the usual generators of the
five-qubit code are $\{ XZZXI, IXZZX, XIXZZ, ZXIXZ \}$ \cite{nc}. For
each qubit $j$, we may pick a stabilizer $S_j$ from this set which acts
as $X$ on that qubit, and $X$ anticommutes with $D_{j} = Z_j$.  This
procedure would be useful in a system where spontaneous emission is the
dominant error process; it would have the virtue of both correcting
spontaneous emission errors by means of feedback as well as correcting
other (rarer) errors by using canonical error correction in addition.

We note that the work in this section can very easily be modified
to generalize the results of \cite{K-Lidar}. That work has the same
error model as ours: known jumps occuring on separate qubits so that the time
and location of each jump is known; but \cite{K-Lidar} postulates fast
unitary pulses instead of a driving Hamiltonian. Their scheme for
spontaneous emission depends on applying the unitary $X^{\otimes n}$ at
intervals $T_c/2$ that are small compared to the rate of spontaneous
emission jumps. They show that after a full $T_c$ period, the no-jump evolution
becomes
\begin{equation}
U = e^{-i T_c H_c / 2} X^{\otimes n} e^{-i T_c H_c / 2 } X^{\otimes n} 
 = e^ { - T_c / 2  \sum_{i = 1}^{N} \kappa_i} 1.
\end{equation}
Thus the application of these pulses acts, as does our driving
Hamiltonian, to correct the no-jump evolution. The generalization from
spontaneous emission to general jump operator $c_j$ for their case is
simple: the code is the same as in the above one-stabilizer protocol,
with single stabilizer equal to (\ref{eqn:oneS}). The fast unitary
pulses are in this case also simply equal to (\ref{eqn:oneS}).

\subsection{Diffusive unraveling}\label{sec:hom}

The case of white-noise feedback, where $\gamma
\rightarrow \infty$, is easily treated by recalling the master 
equation
(\ref{eqn:gen-fb-me}) for white-noise measurement and feedback.  It is
clear that the first term in (\ref{eqn:gen-fb-me}) can be eliminated by
choosing the constant driving Hamiltonian
\begin{equation}
\label{eqn:stab-driv}
H = - (e^{i \phi}c^\dagger F + e^{-i \phi} F c)/2 
\end{equation}
which is automatically Hermitian.
The problem then becomes choosing a feedback 
Hamiltonian $F$ such that $c - i F$
annihilates the codespace. The choice for $F$ can be
made simply by noting that if the codespace is stabilized by some
stabilizer $S$, we can choose
\begin{equation}
\label{eqn:stab-fb}
F = B - i A S.
\end{equation}
Now, note that the decoherence superoperator ${\cal D}$ acts such 
that 
\begin{equation}
{\cal D} [\chi I + L] \rho = {\cal D} [L].
\end{equation}
Then we know that ${\cal D}[c - i F] = {\cal D}[\chi I + A(I - S)]$
annihilates the codespace.

The only caveat is that $F$ is a Hamiltonian and therefore must be
Hermitian. Then the choice (\ref{eqn:stab-fb}) for $F$ is only 
possible
if the anticommutator of $S$ and $A$ is zero:
\begin{equation}
\label{eqn:stab-cond}
\{S,A\} = 0.
\end{equation}
Therefore, if we are given the measurement operator $e^{-i \phi}c = 
\chi
+ A + i B$, we must choose a code with some stabilizer such that
condition (\ref{eqn:stab-cond}) applies; then it is possible to find a
feedback and a driving Hamiltonian such that the total evolution
protects the codespace.

At first glance, it may seem odd that the condition for feedback does
not depend at all upon $B$. This independence has to do with the
measurement unraveling: the diffusive measurement record 
(\ref{eqn:current}) depends only upon  $e^{-i 
\phi}c + e^{i \phi}c^\dagger = 2 (A + \chi)$.

\subsection{Diffusion as the limit of jumps}

It is instructive to show that  the diffusive  
feedback process can be derived by taking the limit of a jump feedback
process using the transformation (\ref{eqn:xform}). 
 This takes several steps, and we use the 
treatment in \cite{wiseman-backaction} as a guide. 
But to begin, note that the condition (\ref{eqn:stab-cond}) 
follows by considering \erf{eqn:ec-gamma} in the limit 
 $\gamma\to\infty$, as the leading order term in $D$ is proportional to 
 $A$.

Consider the jump
unraveling picture with jump operator $c + \gamma$ for $\gamma$ large
(but not infinite). Recall that in the error-correction picture
given in Sec.\ \ref{sec:ham-app}, we postulated a feedback Hamiltonian 
$(dN/dt) V$ that produces a unitary correction
$e^{-iV}$ that acts instantaneously after the jump. In addition we 
will postulate a driving Hamiltonian $K$ that acts when no jump happens. In 
this picture, we will show that given the condition (\ref{eqn:stab-cond}), 
it is possible to find asymptotic expressions for $V$ and $K$ so 
that the deterministic equation for the system
preserves the stabilizer codespace. Finally, we will show that taking 
the limit $\gamma \rightarrow \infty$ leads to the expression for the
feedback and driving Hamiltonians (\ref{eqn:stab-driv}) and
(\ref{eqn:stab-fb}).

Let us consider the measurement operators for the unraveling  with 
large $\gamma$ and $H = 0$. Following
(\ref{eqn:xform}) these are
\begin{eqnarray}
\Omega_1  &=& \sqrt{dt} (c + \gamma)\nonumber\\
\Omega_0 &=& 1 - \frac{dt}{2} [ c \gamma - c^\dagger \gamma  
 + (c + \gamma)^\dagger (c + \gamma)],
\end{eqnarray}
where we have assumed for simplicity that $\gamma$ is real.
Now, including the feedback and driving Hamiltonians modifies these to
\begin{eqnarray}
\label{eqn:prime}
\Omega'_1(dt) &=& \sqrt{dt} e^{-i V} (c + \gamma)\nonumber\\
\Omega'_0(dt) &=& e^{-i K dt} \Omega_0(dt) \nonumber\\
  &=& 1 - i K dt - \frac{dt}{2} (c^\dagger c + 2 \gamma c + \gamma^2).
\end{eqnarray}

Following Ref.~\cite{wiseman-backaction}, 
expand $V$ in terms of $1/\gamma$ to 
second order:
$V = V_1 / \gamma + V_2 / \gamma^2$ where the $V_i$ are Hermitian. 
Then
expanding the exponential in (\ref{eqn:prime}) we get to second order
\begin{eqnarray}
\Omega'_1(dt) &=& \sqrt{dt} \left[1 - i \left(\frac{V_1}{\gamma} +
\frac{V_2}{\gamma^2}\right) - \frac{1}{2} \frac{V_1^2}{\gamma^2}\right] (A + i B 
+ \gamma)
\nonumber\\
&=& \sqrt{dt} \gamma \left[1 +\frac{\chi}{\gamma} + \frac{1}{\gamma} (A + 
i B - i V_1) \right.\nonumber\\
  &&+ \left. \frac{1}{\gamma^2} (V_1^2/2 - i V_2 -i (A + i B) 
  V_1)\right].
\end{eqnarray}

A reasonable choice for $V_1$, by analogy to (\ref{eqn:stab-fb}), is $B
- i A S$. Following \cite{wiseman-backaction}, we also use
(\ref{eqn:stab-driv}) and (\ref{eqn:stab-fb}) to choose $V_2$ and $K$;
note that (\ref{eqn:stab-fb}) is exactly the expression we would expect
for $K$ from (\ref{eqn:gen-ham-driv}) in the limit as $\gamma$ is taken
to infinity.  We will proceed to show that the choice fir $V$ and $K$,
\begin{eqnarray}
\label{eqn:V1}
V_1 &=& B - i A S\\
\label{eqn:V2}
V_2 &=&  - (c^\dagger F + F c)/2\\
\label{eqn:K}
K &=& - \gamma (B - i A S),
\end{eqnarray}
 leads to the
correct evolution to second order in $\gamma$.

Now, the deterministic evolution is given by 
\begin{equation}
\label{eqn:drho}
d \rho =  \Omega_0' \rho \Omega_0' + \Omega_1' \rho \Omega_1' - \rho.
\end{equation}

Substituting (\ref{eqn:prime})--(\ref{eqn:K}) into (\ref{eqn:drho}) to
second order in $\gamma$, after some algebra, gives the deterministic
jump equation
\begin{eqnarray}
\label{eqn:detjump-gamma}
d \rho &=& {\cal D} [A (1 - S)] \rho
\end{eqnarray}
which of course acts as zero on the codespace.

Now we will show that taking the limit as $\gamma
\rightarrow \infty$ leads to the feedback operators given in
(\ref{eqn:stab-driv}) and (\ref{eqn:stab-fb}). We saw in
(\ref{eqn:V1}) and (\ref{eqn:V2}) that the feedback
Hamiltonian needed to undo the effect of the jump operator $c + 
\gamma$
was just 
\begin{equation}
H_{fb} = \frac{dN(t)}{dt} \left(\frac{B - i A S}{\gamma} - \frac{c^\dagger 
F + F c}{2}\right).
\end{equation}
Keeping terms of two orders in $\gamma$ gives
\begin{eqnarray}
H_{fb} &=& \gamma (B - i A S) - \frac{c^\dagger F + F c}{2}\nonumber\\
 && + \frac{dN(t) - \gamma^2 dt}{\gamma dt} (B - i A S).
\end{eqnarray}
The last term just becomes the current $\dot{Q}(t)$ as $\gamma$ 
approaches
infinity, as in equation (\ref{eqn:photocurrent}). Furthermore, we 
have not yet added in the driving Hamiltonian
to the expression for the feedback. Doing so yields
\begin{eqnarray}
H_{\rm total}(t) &=& H_{fb} + K \nonumber\\
 &=& \dot{Q}(t) (B - i A S) - \frac{c^\dagger F + F c}{2}
\end{eqnarray}
which is just what we obtained in the previous section. Thus we can 
see that this continuous current feedback can be thought of as an 
appropriate limit of a jump plus unitary correction process. 

\subsection{Imperfect detection}\label{sec:discussion}
These results for feedback were obtained by assuming unit efficiency,
i.e., perfect detection. Realistically, of course, the efficiency 
$\eta$ will be less than unity. This results in extra terms in the 
feedback master equations we have derived \cite{wiseman_pra94}. 
In the jump case, the extra term is
\beq
\dot{\rho} = (1-\eta) \sum_{j} (c_{j}\rho c_{j}\dg - U_{j}c_{j}\rho 
c_{j}\dg U_{j}\dg).
\eeq
In the diffusion case it is 
\beq
\dot{\rho} = \frac{1 - \eta}{\eta}\sum_{j}{\cal D}[F_{j}].
\eeq
In both cases this results in exponential decay of coherence in the 
codespace. This is because the error correction protocol here relies 
absolutely upon detecting the error when it occurs. If the error is 
missed (jump case), or imperfectly known (diffusion case), then it 
cannot be corrected.  This behavior is, of course, a property of any
continuous-time error correction protocol that depends on correcting
each error instantaneously (e.g., \cite{detected-jump1,detected-jump2,K-Lidar}).

On the other hand, such behavior for Markovian feedback is in
contrast to the state-estimation procedure used in \cite{ADL}. The
latter is much more robust under non-unit efficiency; indeed, given
non-unit efficiency it still works to protect an unknown quantum state
without exponential loss \cite{ADL-qcmc}. This difference in performance
occurs because state-estimation is a function of the entire measurement
record, not just instantaneous measurement results, and thence does not
propagate errors to the same extent that a Markovian feedback system
does. Thus we can see that there is a certain tradeoff. Our Markovian
feedback scheme relies upon calculational simplicity, but at the expense
of robustness. The state-estimation procedure, conversely, is designed
to be robust, but at the cost of computational complexity.

\section{Universal quantum gates}\label{sec:uqc}

Given a protected code subspace, one interesting question, as in
\cite{K-Lidar}, is to investigate what kinds of unitary gates are
possible on such a subspace.  For universal quantum computation on the
subspace--- the ability to build up arbitrary unitary gates on $k$
qubits--- it suffices to be able to perform arbitrary one-qubit gates
for all $k$ encoded qubits and a two-qubit entangling gate such as
controlled-NOT for all encoded qubits $\mu,\nu$. Indeed, as is noted in
\cite{K-Lidar}, it is enough to be able to perform the {\em
Hamiltonians} $\bar{X}_\mu, \bar{Z}_\mu$, and $\bar{X}_\mu \bar{X}_\nu$
for all $\mu,\nu$ \cite{Dodd02}. We will demonstrate that performing
these Hamiltonians with our protocol is possible for the spontaneous
emission scheme given in Sec.\ \ref{sec:ham-app}, and then we will show
how that construction generalizes for an arbitrary jump operator.

Recall that the example in Sec. \ref{sec:ham-app} has single stabilizer
$X^{\otimes n}$ and encodes $n - 1$ logical qubits in $n$ physical
qubits. To find the $2 (n - 1)$ encoded operations, we must find
operators that together with the stabilizer generate the normalizer of
$X^{\otimes n}$ \cite{nc}. In addition, if these operators are to act as
encoded $X$ and $Z$ operations, they must satisfy the usual commutation
relations for these operators:
\begin{eqnarray}
\{X_\mu, Z_\mu \} &=& 0\nonumber\\
\left[X_{\mu}, Z_{\nu}\right] &=& 0, \mu \neq \nu \nonumber\\
\left[X_\mu, X_\nu\right] &=&\left[Z_\mu, Z_\nu\right] = 0.
\end{eqnarray}

Operators satisfying these constraints are easily found for this code:
\begin{eqnarray}
\label{eqn:enc-ops}
\bar{X}_\mu &=& I^{\otimes \mu-1} X I^{\otimes n-\mu} \nonumber\\
\bar{Z}_\mu &=&  I^{\otimes \mu-1} Z I^{\otimes n-\mu-1} Z \nonumber\\
\bar{X}_\mu \bar{X}_\nu &=& I^{\otimes \mu-1} X I^{\otimes \nu-\mu-1} X I^{\otimes n-\nu},
\end{eqnarray}
where we assume $1 \leq \mu < \nu \leq n-1$.
If we apply a Hamiltonian $H_{enc}$ given by any linear combination of
the operators in (\ref{eqn:enc-ops}), the
resulting evolution is encapsulated in the expression for $\Omega_0$,
from (\ref{eqn:omega0-twobit}):
\begin{eqnarray}
\label{eqn:Omega0-gate}
\Omega_0 &=& (1 - \sum_j \kappa_j dt) 1 \nonumber\\
&& - \sum_j \kappa_j Z_j(1 - X^{\otimes n}) dt - i H_{enc} dt.
\end{eqnarray}
As the first term is proportional to
the identity and the second term acts as zero on the codespace, the
effective evolution is given solely by $H_{enc}$ as long as the state
remains in the codespace under that evolution. But because the encoded
operations are elements of the normalizer, as we saw in Sec.\
\ref{bgqecsf}, applying $H_{enc}$ does not take the state out of the
codespace. Furthermore, our protocol assumes that spontaneous emission
jumps are corrected immediately and perfectly, so jumps during the gate
operation will also not take the state out of the codespace. Thus we can
perform universal quantum computation without having to worry about
competing effects from the driving Hamiltonian.

The generalization to the scheme given in Sec. \ref{sec:gen-unrav} to
encode $n-1$ logical qubits in $n$ physical qubits is easily done.
First we note that for the stabilizer $S$
given in the general scheme, we know that
\begin{equation}
S = U X ^ {\otimes n} U^\dagger
\end{equation}
for some unitary $U = \bigotimes_{i=1}^{n} U_i$, so the encoded
operations for that code are similarly given by
\begin{eqnarray}
\bar{X}_\mu &=& I^{\otimes \mu-1}U_\mu X U^\dagger_\mu I^{\otimes n-\mu} \nonumber\\
\bar{Z}_\mu &=& I^{\otimes \mu-1}U_\mu Z U^\dagger_\mu I^{\otimes n-\mu-1}U_n Z U^\dagger_n \nonumber\\
\bar{X}_\mu \bar{X}_\nu 
  &=& I^{\otimes \mu-1}U_\mu X U^\dagger_\mu I^{\otimes \nu-\mu-1}
  U_\nu X U^\dagger_\nu I^{\otimes n-\nu}.
\end{eqnarray}

Now, from (\ref{eqn:gen-ham-driv}) we can see 
that the generalization of (\ref{eqn:Omega0-gate}) is
\begin{equation}
\Omega_0 = (1 - f dt) 1 - \sum_j g_j D_j (1 - S) dt - i H_{enc} dt,
\end{equation}
for real numbers $f, g_i$ given by expanding the expression
(\ref{eqn:omega0}).  Again, since $D(1 - S)$ annihilates the codespace,
the effective evolution is given solely by $H_{enc}$ as long as the
state remains in the codespace under that evolution. Again, $H_{enc}$ is
made up of normalizer elements, which do not take the state out of the
codespace; and again jumps that occur while the gate is being applied
are immediately corrected and thus do not affect the gate operation.
Therefore, universal quantum computation is possible under our general
scheme.\\

\section{Conclusion}\label{sec:conclusion}

We have shown that it is possible to understand a particular variant of
quantum control as quantum error correction. This method is very general
in that it can correct any single qubit detected errors, while requiring
only $n$ physical qubits to encode $n-1$ logical qubits. As a particular
example, we have shown how to correct for spontaneous emission evolution
using feedback and a driving Hamiltonian, which allows less redundancy
than has previously been obtained. We have additionally shown that
universal quantum computation is possible under our method.

We expect that this work will provide a starting point for 
practically implementable  feedback
schemes to protect unknown states. The fact that only two qubits are
required for a demonstration should be particularly appealing. 
We also expect a more complete theoretical development.  
Fruitful avenues for further research include applying notions of
fault-tolerance to this sort of quantum control.

\acknowledgments{C.\ A.\ is grateful for the hospitality of the Centre
for Quantum Computer Technology at the University of Queensland, and
thanks J. Preskill for helpful discussions. C.\ A.\ acknowledges support
from an NSF fellowship and an Institute for Quantum Information
fellowship. This work was supported by the Australian Research Council.}


\begin{thebibliography}{10}

\bibitem{ph219_notes}
J. Preskill, Lecture notes for Caltech course Ph 219: Quantum Information and
  Computation, 1998, http://www.theory.caltech.edu/\verb+~+preskill/ph219/.

\bibitem{nc}
M.~A. Nielsen and I.~L. Chuang, {\em Quantum Computation and Quantum
  Information} (Cambridge University Press, Cambridge, 2000).

\bibitem{secure_QKD}
C.~H. Bennett and G. Brassard,  in {\em Proceedings of IEEE International
  Conference on Computers, Systems and Signal Processing} (PUBLISHER,
  Bangalore, India, 1984), pp.\ 175--179, c.~H. Bennett and G. Brassard,
  ``Quantum public key distribution,'' IBM Technical Disclosure Bulletin {\bf
  28}, 3153--3163 (1985).

\bibitem{shor-ec}
P.~W. Shor, Phys. Rev. A {\bf 52},  2493  (1995).

\bibitem{steane-ec}
A. Steane, Phys. Rev. Lett. {\bf 77},  793  (1996).

\bibitem{knill-laflamme}
E. Knill and R. Laflamme, Phys. Rev. A {\bf 55},  900  (1997),
  quant-ph/9604034.

\bibitem{gott-stab}
D. Gottesman, Phys. Rev. A {\bf 54},  1862  (1996).

\bibitem{wiseman_pra94}
H.~M. Wiseman, Phys. Rev. A {\bf 49},  2133  (1994), errata in Phys. Rev. A
  \textbf{49}, 5159 (1994), Phys. Rev. A \textbf{50}, 4428 (1994).

\bibitem{doherty-jacobs}
A.~C. Doherty and K. Jacobs, Phys. Rev. A {\bf 60},  2700  (1999),
  quant-ph/9812004.

\bibitem{WisManWang02}
H.~M. Wiseman, S. Mancini, and J. Wang, Phys. Rev. A {\bf 66},  013807  (2002).

\bibitem{mabuchi-zoller}
H. Mabuchi and P. Zoller, Phys. Rev. Lett. {\bf 76},  3108  (1996).

\bibitem{qec-spont}
M.~B. Plenio, V. Vedral, and P.~L. Knight, Phys. Rev. A {\bf 55},  67  (1997),
  quant-ph/9603022.

\bibitem{detected-jump1}
G. Alber {\it et~al.}, Phys. Rev. Lett. {\bf 86},  4402  (2001),
  quant-ph/0103042.

\bibitem{detected-jump2}
G. Alber {\it et~al.}, Detected jump-error correcting quantum codes, quantum
  error designs and quantum computation, quant-ph/0208140.

\bibitem{ADL}
C. Ahn, A.~C. Doherty, and A.~J. Landahl, Phys. Rev. A {\bf 65},  042301
  (2002), quant-ph/0110111.

\bibitem{Wineland}
D. Wineland {\it et~al.}, Journal of Research of the National Institute of
  Standards and Technology {\bf 103},  259  (1998).

\bibitem{Vion2002}
D. Vion {\it et~al.}, Science {\bf 296},  886  (2002).

\bibitem{KLM}
E. Knill, R. Laflamme, and G. Milburn, Nature {\bf 409},  46  (2001),
  quant-ph/0006088.

\bibitem{Wiseman2001}
H.~M. Wiseman {\it et~al.}, Phys. Rev. B {\bf 63},  235308  (2001).

\bibitem{doherty_thesis}
A.~C. Doherty, Ph.D. thesis, The University of Auckland, 1999.

\bibitem{K-Lidar}
K. Khodjasteh and D.~A. Lidar, quant-ph/0301105.

\bibitem{ftqc}
J. Preskill,  in {\em Introduction to Quantum Computation and Information},
  edited by H.~K. Lo, S. Popescu, and T. Spiller (World Scientific, New Jersey,
  1998), Chap.~8, p.\ 213, quant-ph/9712048.

\bibitem{wiseman-semiclass}
H.~M. Wiseman, Quantum Semiclassical Optics {\bf 8},  205  (1996).

\bibitem{carmichael}
H.~J. Carmichael, {\em An Open Systems Approach to Quantum Optics}
  (Springer-Verlag, Berlin, 1993).

\bibitem{WallsMilb}
D. F.Walls and G. J.Milburn, {\em Quantum Optics} (Springer-Verlag, Berlin,
  1994).

\bibitem{Milburn1998}
G.~J. Milburn and H.~B. Sun, Phys. Rev. B {\bf 59},  10748  (1999).

\bibitem{fieldquad}
H.~M. Wiseman and G.~J. Milburn, Phys. Rev. A {\bf 47},  642  (1993).

\bibitem{gardiner}
C.~W. Gardiner, {\em Handbook of Stochastic Methods} (Springer, Berlin, 1985).

\bibitem{Goan2001b}
H.-S. Goan, G.~J. Milburn, H.~M. Wiseman, and H.~B. Sun, Physical Review B {\bf
  63},  125326  (2001).

\bibitem{Goan2001}
H.-S. Goan and G.~J. Milburn, Phys. Rev. B {\bf 64},  235307  (2001).

\bibitem{WisMilprl93}
H.~M. Wiseman and G.~J. Milburn, Phys. Rev. Lett. {\bf 70},  548  (1993).

\bibitem{wiseman-backaction}
H.~M. Wiseman, Phys. Rev. A {\bf 51},  2459  (1995).

\bibitem{ADL-qcmc}
C. Ahn, A.~C. Doherty, and A.~J. Landahl,  in {\em Quantum Communication,
  Measurement and Computing (QCMC '02)}, edited by J.~H. Shapiro and O. Hirota
  (Rinton Press, New Jersey, 2003).

\bibitem{Dodd02}
J.~L. Dodd, M.~A. Nielsen, M.~J. Bremner, and R.~T. Thew, Phys. Rev. A {\bf
  65},  040301  (2002).

\end{thebibliography}
\end{document}